\documentclass[runningheads]{llncs}

\usepackage{amssymb}
\usepackage{graphicx}

\newcommand{\be}{\begin{equation}}
\newcommand{\ee}{\end{equation}}
\newcommand{\bra}{\langle}
\newcommand{\ket}{\rangle}
\newcommand{\bea}{\begin{eqnarray}}
\newcommand{\eea}{\end{eqnarray}}
\newcommand{\dis}{\displaystyle}

\newcommand{\keywords}[1]{\par\addvspace\baselineskip
\noindent\keywordname\enspace\ignorespaces#1}

\begin{document}

\title{Financial Time Series Analysis of SV Model by Hybrid Monte Carlo}

\author{Tetsuya Takaishi}

\institute{
         Hiroshima University of Economics,\\
         Hiroshima 731-0192  JAPAN \\
\it         E-mail: {takaishi@hiroshima-u.ac.jp}
}

\maketitle

\begin{abstract}
We apply the hybrid Monte Carlo (HMC) algorithm to 
the financial time sires analysis of  
the stochastic volatility (SV) model for the first time.
The HMC algorithm is used for the Markov chain Monte Carlo (MCMC) update of 
volatility variables of the SV model in the Bayesian inference.
We compute parameters of the SV model from the artificial financial data and 
compare the results from the HMC algorithm with those from the Metropolis algorithm.
We find that the HMC decorrelates the volatility variables
faster than the Metropolis algorithm.
We also make an empirical analysis based on the Yen/Dollar exchange rates.
\end{abstract}

\section{Introduction}
It is well known that 
financial time series of asset returns 
shows various interesting properties which can not be
explained from the assumption of that the time series obeys the Brownian motion.
Those properties are classified as stylized facts\cite{CONT}.
Some examples of the stylized facts are 
(i) fat-tailed distribution of return
(ii) volatility clustering 
(iii) slow decay of the autocorrelation time of the absolute returns.
The true dynamics behind the stylized facts is not fully understood. 
There are some attempts to construct physical models based on spin 
dynamics\cite{SPIN2}-\cite{SPIN6} and they are able to capture some of the stylized facts.   

The volatility is an important value to measure the risk in finance. 
The volatilities of asset returns change in time and shows the volatility clustering.
In order to mimic these empirical properties of the volatility 
Engle advocated the autoregressive conditional hetroskedasticity (ARCH) model\cite{ARCH} 
where the volatility variable changes deterministically depending on the past squared value of the return. 
Later the ARCH model is generalized by 
adding also the past volatility dependence to the volatility change. 
This model is known as the generalized ARCH (GARCH) model\cite{GARCH}.
The parameters of the GARCH model applied to financial time series 
are easily determined by the maximum likelihood method.

The stochastic volatility (SV) model\cite{SVMCMC1,SV} is another model 
which captures the properties of the volatility.  
Contrast to the GARCH model of which the volatility change is deterministic, 
the volatility of the SV model changes stochastically in time.
As a result the likelihood function of the SV model is given as 
a multiple integral of the volatility variables.
Such an integral in general  is not analytically calculable 
and thus the determination of the parameters in the SV model 
by the maximum likelihood method becomes ineffective.

For the SV model the MCMC method based on the Bayesian approach is developed.
In the MCMC of the SV model one has to update the parameter variables and  also the volatility ones. 
The number of the volatility variables to be updated increases  with the data size of the time series.
Usually the update scheme of the volatility variables is based on the local one such as the Metropolis-type 
algorithm\cite{SVMCMC1}. 
It is however known that 
when the local update scheme is done for the volatility variables 
which have interactions to their neighbor variables in time, 
the autocorrelation time of sampled volatility variables becomes high and thus the local update scheme is not effective.
In order to improve the efficiency of the local update method 
the blocked scheme which updates several variables at once is also proposed\cite{SVMCMC2}.

In this paper we use the HMC algorithm\cite{HMC} to 
update the volatility variables.
There exists an application of the HMC algorithm to the GARCH model\cite{HMCGARCH}
where three GARCH parameters are updated by the HMC scheme.
It is more interesting to apply the HMC for update of the volatility variables
because  the HMC algorithm is a global update scheme which can update
all variables at once. 
To examine the HMC we calculate the autocorrelation function
of the volatility variables and compare the result with that of the Metropolis algorithm.

\section{Stochastic Volatility Model and its Bayesian inference}
\subsection {Stochastic Volatility Model}

The standard version of the SV model\cite{SVMCMC1,SV} is given by 
\be
y_t = \sigma_t \epsilon_t = \exp(h_t/2)\epsilon_t,
\label{eq:SV}
\ee
\be
h_t = \mu +\phi (h_{t-1} -\mu) +\eta_t,
\ee
where $y_t=(y_1,y_2,...,y_n)$ represents the time series data, $h_t$ is defined 
by $h_t=\ln \sigma_t^2$ 
and $\sigma_t$ is called volatility.
The error terms $\epsilon_t$ and $\eta_t$ are independent normal distributions
$N(0,1)$ and $N(0,\sigma_\eta^2)$ respectively.

For this model the parameters to be determined are $\mu$, $\phi$ and $\sigma^2_\eta$.
Let us use $\theta$ as  $\theta=(\mu, \phi,\sigma^2_\eta)$.
The likelihood function $L({\bf \theta})$ for the SV model is written as 
\be
L({\bf \theta})=\int \prod_{t=1}^n f(\epsilon_t|\sigma_t^2) f(h_t|\theta) dh_1dh_2...dh_n,
\label{LFUNC}
\ee
where
\be
f(\epsilon_t|\sigma_t^2)=(2\pi \sigma_t^2)^{-\frac12}\exp(-\frac{y_t^2}{2\sigma_t^2}),
\ee
\be
f(h_1|\theta)=(\frac{2\pi \sigma_\eta^2}{1-\phi^2})^{-\frac12}  
\exp(-\frac{[h_1-\mu]^2}{2\sigma_\eta^2/(1-\phi^2)}),
\ee

\be
f(h_t|\theta)=(2\pi\sigma_\eta^2)^{-\frac12} 
\exp(-\frac{[h_t-\mu-\phi(h_{t-1}-\mu)]^2 }{2\sigma_\eta^2}).
\ee

\subsection{Bayesian inference for the SV model}
In the Bayesian theorem, the probability distributions of the parameters to be estimated
are given by
\be
f(\theta|y)=\frac1Z L({\bf \theta}) \pi(\bf \theta),
\ee
where $Z$ is the normalization constant $Z=\int L({\bf \theta}) \pi({\bf \theta}) d\theta$ and
$\pi(\bf \theta)$ is a prior distibution of ${\bf \theta}$ for which we make a certian assumption.
The values of the parameters are inferred as the expectation values of  
$\theta$ given by 
$\bra {\bf \theta} \ket = \int {\bf \theta} f(\theta|y) d\theta$.
In general this integral can not be performed analytically.
For that case, one can use the MCMC method to estimate the expectation values numerically. 
In the MCMC method, we first generate a series of $\theta$  with 
a probability $P(\theta)= f(\theta|y)$.
Let $\theta^{(i)}=(\theta^{(1)},\theta^{(2)},...,\theta^{(k)})$ be values of $\theta$ generated  
by a MCMC sampling. Then using these values the expectation value of $\theta$ is estimated by
$\bra {\bf \theta} \ket = \frac1k\sum_{i=1}^k \theta^{(i)}.
$

For the SV model, in addition to $\theta$, volatility variables $h_t$ also
have to be updated since they are integrated out as in eq.(\ref{LFUNC}). 
Let $P(\theta,h_t)$ be the joint probability distribution of
$\theta$ and $h_t$.
Then $P(\theta,h_t)$ is given by
\be 
P(\theta,h_t) \sim \bar{L}(\theta,h_t)\pi(\theta),
\label{eq:prob}
\ee
where 
$\bar{L}(\theta,h_t) = \prod_{t=1}^n f(\epsilon_t|h_t) f(h_t|\theta)$.

For the prior $\pi(\theta)$ we assume that $\pi(\sigma_\eta^2)\sim (\sigma_\eta^2)^{-1}$
and for others $\pi(\mu) =\pi(\phi) =constant.$
The probability distributions for the parameters and the volatility variables
are derived from eq.(\ref{eq:prob})\cite{SVMCMC1,SV}.
The probability distributions and their update schemes  
are given in the followings.

\begin{itemize}
\item $\sigma_\eta^2$ update scheme.

The probability distribution of $\sigma_\eta^2$ is given by
\be
P(\sigma_\eta^2)\sim (\sigma_\eta^2)^{-\frac{n}2-1} \exp\left(-\frac{A}{\sigma_\eta^2}\right),
\label{sigma}
\ee
where $A=\frac12\{(1-\phi^2)(h_1-\mu)^2+\sum_{t=2}^n [h_t-\mu -\phi(h_{t-1}-\mu)]^2\}$.

Since eq.(\ref{sigma}) is an inverse gamma distribution 
we can easily draw a value depending on eq.(\ref{sigma})
by using an appropriate statistical library.

\item $\mu$ update scheme.

The probability distribution of $\mu$ is given by
\be
P(\mu) \sim \exp\left\{-\frac{B}{2\sigma_\eta^2}(\mu-\frac{C}{B})^2 \right\},
\label{mu}
\ee
where $ B=(1-\phi^2)+(n-1)(1-\phi)^2$, \\
and $C=(1-\phi^2)h_1+(1-\phi)\sum_{t=2}^n (h_t-\phi h_{t-1})$.

Eq.(\ref{mu}) is a Gaussian distribution. Again we can easily update $\mu$.

\item $\phi$ update scheme.

The probability distribution of $\phi$ is given by
\be
P(\phi)\sim (1-\phi^2)^{1/2}\exp\{-\frac{D}{2\sigma_\eta^2}(\phi-\frac{E}{D})^2\},
\label{phi}
\ee
where $D=-(h_1-\mu)^2+\sum_{t=2}^n(h_{t-1}-\mu)^2$ and $E=\sum_{t=1}^n (h_t-\mu)(h_{t-1}-\mu)$.

In order to update $\phi$ with eq.(\ref{phi}), we use the Metropolis-Hastings algorithm\cite{METRO,MH}.
Let us write eq.(\ref{phi}) as
$ \dis P(\phi)\sim P_1(\phi)P_2(\phi)$ where
\be
P_1(\phi)= (1-\phi^2)^{1/2},
\ee
\be
P_2(\phi) \sim \exp\{-\frac{D}{2\sigma_\eta^2}(\phi-\frac{E}{D})^2\}.
\label{phi2}
\ee
Since $P_2(\phi)$ is a Gaussian distribution we can easily draw $\phi$ from eq.(\ref{phi2}).
Let $\phi_{new}$ be a candidate given from eq.(\ref{phi2}).
Then in order to obtain the correct distribution,
$\phi_{new}$ is accepted with the following probability $P_{MH}$.
\be
P_{MH}=\min\left\{\frac{P(\phi_{new})P_2(\phi)}{P(\phi)P_2(\phi_{new})},1 \right\}
=\min\left\{\sqrt{\frac{(1-\phi^2_{new})}{(1-\phi^2)}},1\right\}.
\ee
In addition to the above step we restrict $\phi$ within $[-1,1]$ to avoid a negative value 
in the calculation of square root.

\item Probability distribution for $P(h_t)$.

The probability distribution of the volatility variables $h_t$ 
is given by   
\bea 
\label{eq:ham}
& P(h_1,h_2,...,h_n)  \sim \vspace{2cm}  \\ \nonumber
& \exp \left(-\sum_{i=1}^n \{\frac{h_t}{2}+\frac{\epsilon_t^2}{2}e^{-h_t}\}
\frac{[h_1-\mu]^2}{2\sigma_\eta^2/(1-\phi^2)}
-\sum_{i=2}^n \frac{[h_t-\mu-\phi(h_{t-1}-\mu)]^2}{2\sigma_\eta^2}\right).
%\label{eq:ham}
\eea
This probability distribution is not a simple function 
for drawing values of the volatility variables $h_t$.
A conventional method is the Metropolis method\cite{METRO} which updates the variables
locally. 
Here we use the HMC algorithm which updates the volatility variables globally.

\end{itemize}

\section{Hybrid Monte Carlo Algorithm}
The HMC algorithm is originally developed for the MCMC
of the lattice Quantum Chromo Dynamics (QCD) calculations \cite{HMC}
where local type update algorithms are not effective. 
The notable feature of the HMC algorithm is that
it updates a number of variables simultaneously.

Here we briefly describe the HMC algorithm.
The HMC algorithm combines molecular dynamics (MD) simulations 
and the Metropolis test.
Let $f(x)$ be  a probability density and $O(x)$  a function of $x=(x_1,x_2,...,x_n)$.
We determine the expectation value of $O(x)$ with the probability density $f(x)$
which is given by 
\be
\bra O(x) \ket   =  \int O(x) f(x) dx   =  \int O(x) e^{ln f(x)} dx.
\label{eq:integral}
\ee

Now let us introduce momentum variables $p=(p_1,p_2,...,p_n)$ conjugate to the variables $x$  and 
rewrite eq.(\ref{eq:integral}) as
\be 
\bra O(x) \ket   =   \frac1{Z}  \int O(x) e^{- \frac12 p^2 + ln f(x)} dxdp  =   \frac1{Z}  \int O(x) e^{- H(p,x)} dxdp.
\ee
where $Z=\int  e^{- \frac12 p^2 } dp$.  
$H(p,x)$ is the Hamiltonian defined by $H(p,x) = \frac12 p^2 - ln f(x)$
where $p^2$ stands for $\sum_{i=1}^n p_i^2$.
The introduction of $p$ does not change the value of $\bra O(x) \ket$.

In the HMC algorithm, new candidates of the variables are drawn by 
integrating the Hamilton's equations of motions.
The Hamilton's equations of motions are solved numerically by doing the MD simulation with a fictitious time.
To solve the equations we use the standard 2nd order leapfrog integrator.
One could use improved integrators\cite{MNHOMC} 
or higher order integrators\cite{HOHMC,HOHMC2} if necessary.

Let $(p',x')$ be the new candidates given by the MD simulation.
The new candidates are accepted with a probability $\min\{1,\exp(-\Delta H)\}$
where $\Delta H = H(p',x')-H(p,x)$.
Since the Hamilton's equation of motions are not solved exactly 
$\Delta H$ deviates from zero.
The magnitude of the deviation is tuned by the discrete time step size  
in the MD simulation such that the acceptance of the new candidates becomes high. 

For the volatility variables of the SV model, from eq.(\ref{eq:ham}) 
the Hamiltonian can be defined by
\be
H(p_t,h_t)=\sum_{i=1}^n \frac12 p_i^2 +
\sum_{i=1}^n \{\frac{h_i}{2}+\frac{\epsilon_i^2}{2}e^{-h_i}\}
+\frac{[h_1-\mu]^2}{2\sigma_\eta^2/(1-\phi^2)}
+\sum_{i=2}^n \frac{[h_i-\mu-\phi(h_{i-1}-\mu)]^2}{2\sigma_\eta^2},
\ee
where $p_i$ is defined as a conjugate momentum to $h_i$.

\section{Numerical Test}
In this section we investigate the HMC algorithm for the SV model 
with artificial financial data.
The artificial data is generated with a set of known parameters.
We try to infer the values of those parameters by the HMC and 
Metropolis algorithms and compares the results. 

Using eq.(\ref{eq:SV}) with $\phi=0.97$,$\sigma_{\eta}^2 = 0.05$ and $\mu=-1$ 
we have generated 2000 data. 
To this data we made the Bayesian inference with the HMC
and Metropolis algorithms. 
The initial parameters are set to $\phi=0.5$,$\sigma_{\eta}^2 = 1.0$ and $\mu=0$.
The first 10000 samples are discarded as thermalization or burn-in process.
Then 200000 samples are recorded for analysis. The acceptance of the volatility variables is 
tuned to be about 50\%.

\begin{figure}
\vspace{2mm}
\centering
\includegraphics[height=4.2cm]{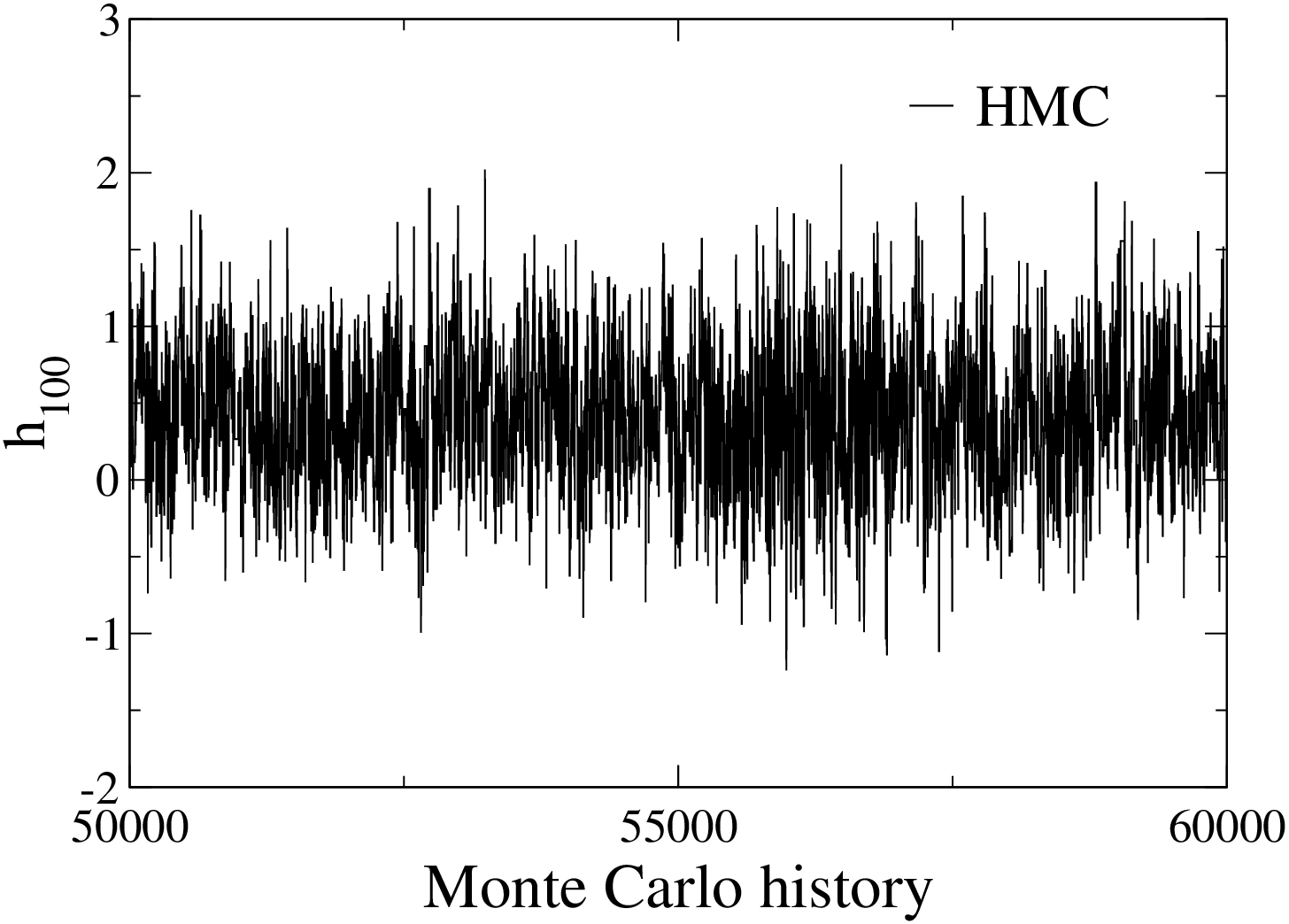}
\hspace{2mm}
\includegraphics[height=4.2cm]{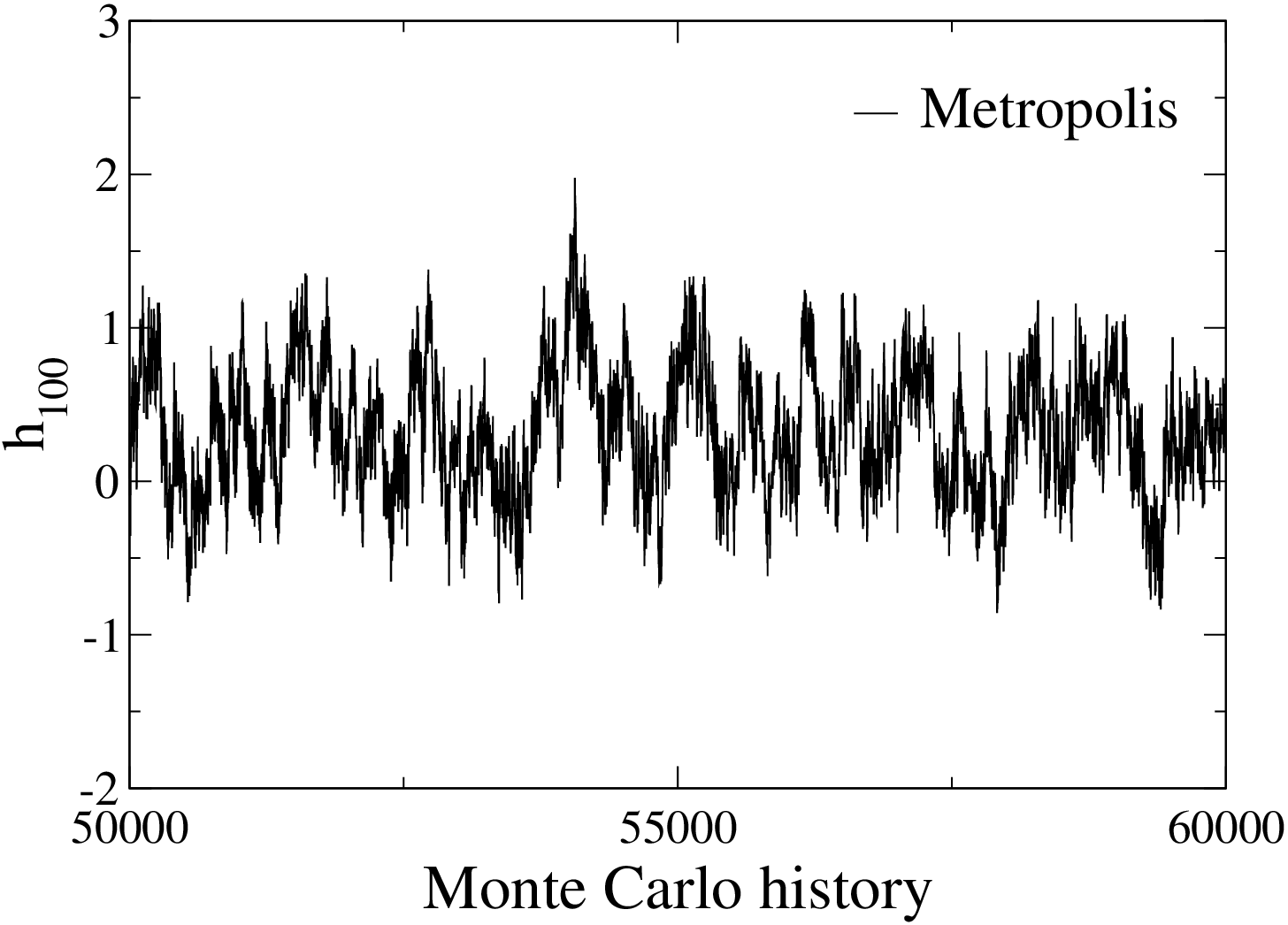}
\vspace{-1mm}
\caption{
Monte Carlo history of HMC (left) and Metropolis (right)
in the window from 50000 to 60000 Monte Carlo history.}
\label{fig:traj}
\vspace{-1mm}
\end{figure}

\begin{figure}
%\vspace{1mm}
\centering
\includegraphics[height=4.1cm]{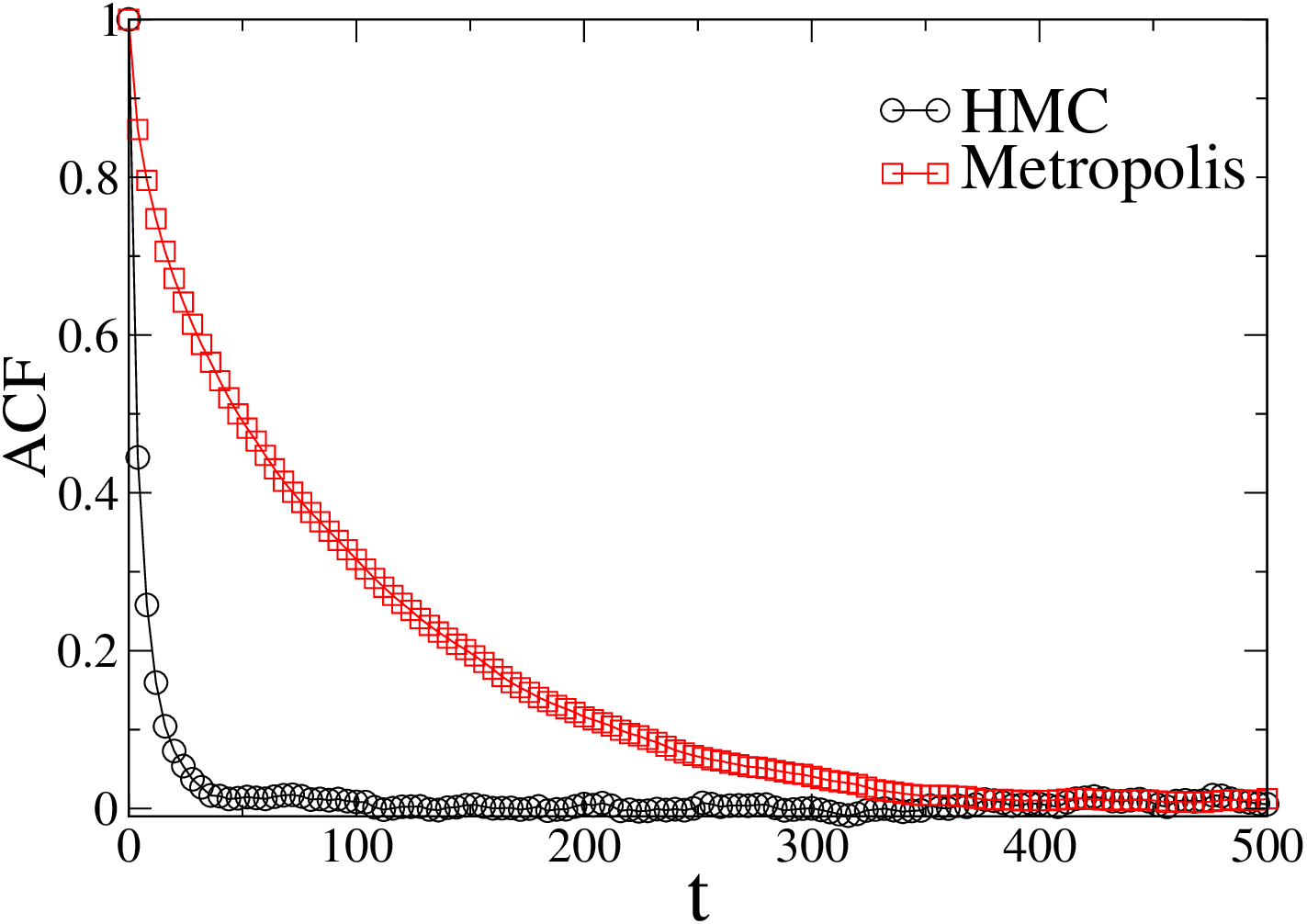}
%\hspace{2mm}
%\includegraphics[height=4.0cm]{hist_phiall.eps}
\vspace{-2mm}
\caption{
Autocorrelation function of the volatility variable $h_{100}$
for the HMC and Metropolis algorithms. 
We see that the ACF from HMC decreases quickly as the Monte Carlo time increases.}
\label{fig:auto}
\vspace{-1mm}
\end{figure}

Fig.\ref{fig:traj} shows the history of the volatility variable $h_{100}$. 
We use $h_{100}$ as the representative one of the volatility variable. 
We observe the similar behavior for other volatility variables.
As seen in Fig.\ref{fig:traj}
the correlation of the volatility variable from the HMC algorithm is smaller than 
that from the Metropolis algorithm. 
To quantify this we calculated the autocorrelation function (ACF) of the volatility variable
shown in Fig.\ref{fig:auto}.
The ACF is defined as 
\be
ACF(t) = \frac{\frac1N\sum_{j=1}^N(x(j)- <x> )(x(j+t)-<x>)}{\sigma^2_x},
\ee
where $<x>$ and $\sigma^2_x$ are the average value and the variance of $x$ respectively.

The autocorrelation time $\tau_{int}$ of the volatility variables is given in Table 1.
The values in the parentheses represent the errors estimated by the jackknife method.
The autocorrelation time is defined by
$\tau_{int} = \frac12 + \sum_{t=1}^{\infty} ACF(t)$.

The HMC algorithm gives a smaller autocorrelation time than the Metropolis algorithm,
which means that the HMC algorithm samples the volatility variables more effectively 
than the Metropolis algorithm.

\begin{table}
\centering
\begin{tabular}{c|cccc}
             & \makebox[20mm]{$\phi$}   & \makebox[20mm]{$\mu$}  &  \makebox[20mm]{$\sigma^2_{\eta}$}   &  \makebox[20mm]{$h_{100}$}    \\  \hline
true         & 0.97                     &  -1                    &   0.05                               &                            \\ \hline
HMC          &  0.978(7)   & -0.92(26)  &  0.053(12)         &                     \\
$\tau_{int}$ &  540(60)    & 3(1)       &  1200(150)         &   18(1)             \\ \hline
Metropolis   &  0.978(7)   & -0.92(26)  &  0.052(11)         &                     \\
$\tau_{int}$ & 400(100)    & 13(2)      &  1000(270)         &   210(50)
\end{tabular}
\caption{Results estimated by the HMC and Metropolis algorithms.}
\vspace{-2mm}
\end{table}

\begin{table}
\vspace{-1mm}
\centering
\begin{tabular}{c|cccc}
         &  \makebox[20mm]{$\phi$}   & \makebox[20mm]{$\mu$}  &  \makebox[20mm]{$\sigma^2_{\eta}$}   &  \makebox[20mm]{$h_{100}$}    \\  \hline
HMC          &  0.960(12)   & -1.13(8)  &  0.014(4)         &                     \\
$\tau_{int}$ &  610(300)    & 14(6)       &  1400(800)         &   55(11)             \\ 
\end{tabular}
\caption{Results estimated by the HMC to the Yen/Dollar exchange rates.}
\vspace{-2mm}
\end{table}

The autocorrelation times for the parameters of the SV model are summarized in Table 1.
The autocorrelation times from the HMC algorithm are similar to those of the Metropolis algorithm except for $\tau_{int}$ of $\mu$. 

The values of the SV parameters estimated by the HMC and the Metropolis algorithms are given in Table 1.
The values in the parentheses represent the standard deviations of the sampled data. 
The results from the both algorithms well reproduce the true values used for the generation of the artificial financial data.
Furthermore for each parameter two values obtained by the HMC and the Metropolis algorithms agree well.
This is not surprising because the same data is used for the both calculations by the HMC and Metropolis algorithms.

\section{Empirical Study}
We have also made an empirical study of the SV model by the HMC.
The empirical study is based on daily data of the exchange rates for Japanese Yen and US dollar.
The sampling period is  1 March 2000  to 29 February 2008, which has 2007 observations.
The exchange rates $p_i$ are transformed to $r_i=100\ln(p_i/p_{i-1}-\bar{s})$ 
where $\bar{s}$ is the average value of $\ln(p_i/p_{i-1})$. 
The MCMC sampling is performed as in the previous section.
The first 10000 MC samples are discarded and then 20000 samples are recoded for the analysis. 
The estimated values of the parameters are summarized in Table 2.
The estimated value of $\phi$ is close to one, which means the persistency of 
the volatility shock. The similar values are obtained in the previous studies\cite{SVMCMC1,SV}.

\section{Summary}
The HMC algorithm is applied for the Bayesian inference of the SV model.
It is found that the correlations of the volatility variables sampled by the HMC algorithm are much reduced.
On the other hand we observe no significant improvement on the correlations of the sampled parameters of the SV model.
Thus it is concluded that the HMC algorithm has a similar efficiency to the Metropolis algorithm  
and it is an alternative algorithm for the Bayesian inference of the SV model. 

If one needs to calculate a certain quantity depending on the volatility variables,
then the HMC algorithm may serve as a good algorithm which samples 
the volatility variables effectively because the HMC algorithm decorrelates 
the sampled volatility variables faster than the Metropolis algorithm. 

\subsubsection*{Acknowledgments.}
The numerical calculations were carried out on SX8 at the Yukawa Institute for Theoretical Physics in Kyoto University
and on Altix at the Institute of Statistical Mathematics.
The author thanks K.Maekawa, Y.Tokutsu, T.Morimoto, K.Kawai, K.Tei and X.H. Lu
for valuable discussions.


\begin{thebibliography}{99}
\bibitem{CONT}
See e.g., Cont, R.: Empirical Properties of Asset Returns: Stylized Facts and Statistical Issues.
Quantitative Finance 1, 223-236  (2001)

\bibitem{SPIN2}
Bornholdt, S.: 
Expectation Bubbles in a Spin Model of Markets: Intermittency from Frustration across Scales.
%EXPECTATION BUBBLES IN A SPIN MODEL OF MARKETS: INTERMITTENCY FROM FRUSTRATION ACROSS SCALES
Int. J. Mod. Phys. C 12, 667 - 674 (2001)

\bibitem{SPIN4}
Sznajd-Weron, K., Weron, R.:
A Simple Model of Price Formation.
%A SIMPLE MODEL OF PRICE FORMATION
Int. J. Mod. Phys. C 13, 115 - 123 (2002)

\bibitem{SPIN5}
Kaizoji, T. {\it et al.}
%, Bornholdt, S., Fujiwara, Y.:
: Dynamics of Price and Trading Volume in a Spin Model of Stock Markets with Heterogeneous Agents.
Physica A 316, 441-452 (2002)

\bibitem{SPIN6}
Takaishi, T.:
Simulations of Financial Markets in a Potts-like Model 
Int. J. Mod. Phys. C 13, 1311 - 1317 (2005) 

\bibitem{ARCH}
Engle, R.F.: Autoregressive Conditional Heteroskedasticity with Estimates of the Variance.
of the United Kingdom inflation.
Econometrica 60, 987-1007 (1982) 

\bibitem{GARCH}
Bollerslev, T.: Generalized Autoregressive Conditional Heteroskedasticity.
Journal of Econometrics 31, 307-327 (1986) 

\bibitem{SVMCMC1}
Jacquier, E., Polson, N.G., Rossi, P.E.: Bayesian Analysis of Stochastic Volatility Models.
Journal of Business \& Economic Statistics, 12 (1994) 371

\bibitem{SV}
Kim, S., Shephard, N., Chib, S.: Stochastic Volatility: Likelihood Inference and Comparison with ARCH Models.
Review of Economic Studies 65, 361-393 (1998)

\bibitem{SVMCMC2}
Shephard, N., Pitt, M.K.: Likelihood Analysis of Non-Gaussian Measurement Time Series.
Biometrika 84, 653-667 (1994)

\bibitem{HMC}
Duane, S. {\it et al.}
%Kennedy, A.D, Pendleton, B.J., Roweth, D.
: Hybrid Monte Carlo.
Phys. Lett. B 195, 216-222 (1987)

\bibitem{HMCGARCH}
Takaishi, T.: Bayesian Estimation of GARCH model by Hybrid Monte Carlo.
Proceedings of the 9th Joint Conference on Information Sciences 2006, CIEF-214 \\
doi:10.2991/jcis.2006.159

\bibitem{METRO}
Metropolis, N. {\it et al.}
%, Rosenblut, A.W., Rosenbluth, M.N., Teller, A.H., Teller, E.
: Equations of State Calculations by Fast Computing Machines.
J. of Chem. Phys. 21, 1087-1091 (1953)

\bibitem{MH}
Hastings, W.K.: Monte Carlo Sampling Methods Using Markov Chains and Their Applications.
Biometrika 57, 97-109 (1970)


\bibitem{MNHOMC}
Takaishi, T., de Forcrand, Ph.:
Testing and Tuning Symplectic Integrators for Hybrid Monte Carlo Algorithm in Lattice QCD.
Phys. Rev. E 73, 036706 (2006)

\bibitem{HOHMC}
Takaishi, T: Choice of Integrators in the Hybrid Monte Carlo Algorithm.
Comput. Phys. Commun. 133, 6-17 (2000) 

\bibitem{HOHMC2}
Takaishi, T: Higher Order Hybrid Monte Carlo at Finite Temperature.
Phys. Lett. B 540, 159-165 (2002) 

\end{thebibliography}
\end{document}